\documentclass[twocolumn]{aastex62}
\usepackage{natbib}
\usepackage{epsfig}
\usepackage{graphicx}
\usepackage{subfigure}
\usepackage{float}
\usepackage{amsmath}
\usepackage{color}
\usepackage{amssymb}
\usepackage{amsfonts}
\usepackage{apjfonts}
\newcommand{\PMO}{Purple Mountain Observatory, Chinese Academy of Sciences, Nanjing 210023, China}
\newcommand{\GXU}{Guangxi Key Laboratory for Relativistic Astrophysics, Guangxi University, Nanning 530004, China}
\newcommand{\UCAS}{University of Chinese Academy of Sciences, Beijing 100049, China}
\newcommand{\UA}{Department of Physics, The Applied Math Program, and Department of Astronomy, The University of Arizona, Tucson, AZ 85721, USA}

\shortauthors{Wei \& Melia}

\begin{document}

\title{Cosmology-independent Estimate of the Hubble Constant and Spatial Curvature Using Time-delay Lenses and Quasars}
%$H_0$ and $\Omega_{K}$

\correspondingauthor{Jun-Jie Wei}
\email{jjwei@pmo.ac.cn}

\author{Jun-Jie Wei}
\affiliation{\PMO}
\affiliation{\GXU}
\affiliation{\UCAS}

\author{Fulvio Melia \thanks{John Woodruff Simpson Fellow.}}
\affiliation{\UA}

\begin{abstract}
With the distance sum rule in the Friedmann-Lema\^{\i}tre-Robertson-Walker metric, model-independent
constraints on both the Hubble constant $H_0$ and spatial curvature $\Omega_{K}$ can be obtained using
strong lensing time-delay data and Type Ia supernova (SN Ia) luminosity distances. This method is
limited by the relative low redshifts of SNe Ia, however. Here, we propose using quasars as distance indicators,
extending the coverage to encompass the redshift range of strong lensing systems. We provide a novel and
improved method of determining $H_0$ and $\Omega_{K}$ simultaneously. By applying this technique to
the time-delay measurements of seven strong lensing systems and the known ultraviolet versus X-ray luminosity
correlation of quasars, we constrain the possible values of both $H_0$ and $\Omega_{K}$, and find that
$H_0=75.3^{+3.0}_{-2.9}$ km $\rm s^{-1}$ $\rm Mpc^{-1}$ and $\Omega_{K}=-0.01^{+0.18}_{-0.17}$.
The measured $\Omega_{K}$ is consistent with zero spatial curvature, indicating that there is no
significant deviation from a flat universe. If we use flatness as a prior, we infer that
$H_0=75.3^{+1.9}_{-1.9}$ km $\rm s^{-1}$ $\rm Mpc^{-1}$, representing a precision of
2.5\%. If we further combine these data with the 1048 current Pantheon SNe Ia, our
model-independent constraints can be further improved to $H_0=75.3^{+3.0}_{-2.9}$ km
$\rm s^{-1}$ $\rm Mpc^{-1}$ and $\Omega_{K}=0.05^{+0.16}_{-0.14}$. In every case, we find
that the Hubble constant measured with this technique is strongly consistent with the value
($\sim 74$ km $\rm s^{-1}$ $\rm Mpc^{-1}$) measured using the local distance ladder, as opposed
to the value optimized by {\it Planck}.
\end{abstract}


\section{Introduction}

The Hubble constant $H_0$ characterizes the current expansion rate of the Universe and determines its
absolute distance scale. In recent years, the accuracy of measuring $H_0$ has been significantly
improved, but the value of $H_0$ ($=67.4\pm0.5$ km $\rm s^{-1}$ $\rm Mpc^{-1}$;
\citealt{2018arXiv180706209P}) inferred from {\it Planck} observations of the cosmic microwave background (CMB)
in the context of flat $\Lambda$CDM represents a $4.4\sigma$ tension with that measured from
local type Ia supernovae (SNe Ia) calibrated by the Cepheid distance ladder ($H_0=74.03\pm1.42$
km $\rm s^{-1}$ $\rm Mpc^{-1}$; \citealt{2019ApJ...876...85R}). Other early-Universe probes,
such as a combination of clustering and weak lensing, baryon acoustic oscillations, and big-bang
nucleosynthesis, yield results similar to the CMB \citep{2018MNRAS.480.3879A}, while an alternate
local calibration of the distance ladder using the tip of the red giant branch finds an intermediate
value of $H_0$ (\citealt{2019ApJ...882...34F,2020ApJ...891...57F}; but see
\citealt{2019ApJ...886...61Y}). A review for the current status of the Hubble tension may be found
in \cite{2019NatAs...3..891V}. If the systematic errors of the observations cannot account for the
discrepancy, this Hubble tension may indicate new physics beyond the standard $\Lambda$CDM
cosmological model.

To better understand the origin of the tension, more independent determinations of $H_0$ are required.
Strong gravitational lensing provides an independent method of measuring $H_0$
\citep{1964MNRAS.128..307R}. The time delay between strongly lensed images of variable sources is
related to a quantity called the ``time-delay distance'', $D_{\Delta t}$, which depends on the lensing
potential. The quantity $D_{\Delta t}$ is a ratio of three angular diameter distances between the
observer, lens, and source, and is primarily sensitive to $H_0$, but also weakly dependent on other
cosmological parameters. Thus, the Hubble constant $H_0$ can be constrained by these sources. This
method is completely independent of, and complementary to, both the CMB and distance ladder
analyses. Measuring $H_0$ in this manner, however, one has to assume a background cosmology.
Recently, the $H_0$ Lenses in COSMOGRAIL's Wellspring (H0LiCOW) collaboration derived
$H_0=73.3^{+1.7}_{-1.8}$ km $\rm s^{-1}$ $\rm Mpc^{-1}$
for flat $\Lambda$CDM using a sample of six gravitational lens time delays. The value of $H_0$
changed to $81.6^{+4.9}_{-5.3}$ km $\rm s^{-1}$ $\rm Mpc^{-1}$, however, for flat $w$CDM, in which
the dark-energy equation of state is not fixed to $-1$ a priori \citep{2019arXiv190704869W}. Obviously,
the inferred value of $H_0$ using time-delay cosmography is strongly model dependent.

Instead of computing the time-delay distances within a specific cosmological model, one can determine
the angular diameter distances from the observer to the source and lens through observations of SNe Ia
to obtain model-independent constraints on $H_0$ (e.g., \citealt{2015PhRvD..92l3516A,2015MNRAS.448.3463C,2019PhRvL.123w1101C,2019ApJ...886L..23L,2020arXiv200210605L,2019arXiv191204325P}).
But the relation of these two distances and the angular diameter distance from the lens to the source
can not be determined directly from the observations. These three distances in the
Friedmann-Lema\^{\i}tre-Robertson-Walker (FLRW) metric are connected via the distance sum rule,
which depends on the curvature parameter of the Universe. In turn, under the assumption that the Universe
is described by the FLRW metric, both $H_0$ and the spatial curvature can be estimated independently
of the model by combining observations of strong lensing and SNe Ia \citep{2019PhRvL.123w1101C}.
Furthermore, the comparision of the inferred values of the cosmic curvature from two or more lens-source
pairs provides a consistency test of the FLRW metric \citep{2015PhRvL.115j1301R}. Based on the sum rule
of distances along null geodesics of the FLRW metric, model-independent determinations of the spatial
curvature have been implemented by combining strong gravitational lensing systems with other distance
indicators, including SNe Ia \citep{2015PhRvL.115j1301R,2017ApJ...839...70L,2019ApJ...886L..23L,2020arXiv200210605L,2017ApJ...834...75X,2018JCAP...03..041D,2018ApJ...854..146L,
2019PhRvL.123w1101C,2020ApJ...889..186Z}, gravitational waves (GWs; \citealt{2019PhRvD..99h3514L}),
and compact radio sources \citep{2019MNRAS.483.1104Q}. Such model-independent curvature determinations
have also been proposed using future time delay measurements of strongly lensed transients (such as
fast radio bursts, GWs, and SNe) and luminosity distances of SNe Ia \citep{2018NatCo...9.3833L,2019ApJ...873...37L,2019PhRvD.100b3530Q}.

Among these studies, \cite{2019PhRvL.123w1101C} was the first to apply such a method to real data. They
used combined observations of strong lensing time delays and SN Ia luminosity distances to determine not
only the spatial curvature but also $H_0$ without adopting any particular model (see also
\citealt{2019ApJ...886L..23L,2020arXiv200210605L}). It must be emphasized, however, that SNe Ia may be
seen only up to $z\sim2$, while the redshifts of the lens sources detected by the Large Synoptic Survey
Telescope (LSST) would reach $z\sim5$ \citep{2017ApJ...839...70L}. We can therefore only employ a small
fraction of the lensing data that overlaps with the observed SNe Ia for this analysis. Using the
distance sum rule would benefit considerably from the use of other distance indicators extending to
higher redshifts, thus taking full advantage of the whole lensing catalog. Thanks to their high luminosities,
quasars have been viewed as promising cosmological probes. One can estimate their luminosity distances based
on a nonlinear correlation between their ultraviolet (UV) and X-ray monochromatic luminosities. Although
this correlation has been known for more than 30 years \citep{1986ApJ...305...83A}, only recently has the
uncomfortably large dispersion in the relation been mitigated by refining the selection technique and
flux measurements \citep{2015ApJ...815...33R,2019NatAs...3..272R,2016ApJ...819..154L,2017A&A...602A..79L}.
This offers the possibility of using quasars as distance indicators, extending to redshifts $\sim6$. In
this paper, we propose to use the wide redshift coverage of quasars to fully exploit sample of strong
lensing systems in the LSST, in order to simultaneously measure $H_0$ and the spatial curvature, with
the hope of providing more stringent constraints. In this paper, we use the updated H0LiCOW and
STRong-lensing Insights into Dark Energy Survey (STRIDES) dataset consisting of seven lenses \citep{2019arXiv190704869W,2020MNRAS.tmp.1051S} in order to extract the time-delay distances,
and use the recently compiled, high-quality catalogue of 1598 UV and X-ray flux measurements of quasars
covering the redshift range $0.035<z<5.1$ \citep{2019NatAs...3..272R} to obtain the distance-redshift
relation.

\begin{table*}
\centering \caption{Redshifts and time-delay distances for the six H0LiCOW lenses and one STRIDES lens}
\begin{tabular}{lllll}
\hline
\hline
 Lens name &  $z_l$  &  $z_s$  &  $D_{\Delta t}$ (Mpc)  & References \\
\hline
B1608+656      &   0.6304      &   1.394   &   $5156^{+296}_{-236}$  &  \cite{2010ApJ...711..201S,2019Sci...365.1134J}   \\
RXJ1131-1231   &   0.295       &   0.654   &   $2096^{+98}_{-83}$    &  \cite{2014ApJ...788L..35S,2019MNRAS.490.1743C}   \\
HE 0435-1223   &   0.4546      &   1.693   &   $2707^{+183}_{-168}$  &  \cite{2017MNRAS.465.4895W,2019MNRAS.490.1743C}   \\
SDSS 1206+4332 &   0.745       &   1.789   &   $5769^{+589}_{-471}$  &  \cite{2019MNRAS.484.4726B}   \\
WFI2033-4723   &   0.6575      &   1.662   &   $4784^{+399}_{-248}$  &  \cite{2019arXiv190509338R}   \\
PG 1115+080    &   0.311       &   1.722   &   $1470^{+130}_{-127}$  &  \cite{2019MNRAS.490.1743C}   \\
\hline
DES J0408-5354 &   0.597       &   2.375   &   $3382^{+146}_{-115}$  &  \cite{2017ApJ...838L..15L,2020MNRAS.tmp.1051S}   \\
\hline
\end{tabular}
\label{table1}
\end{table*}

The outline of this paper is as follows. In Section~\ref{sec:method}, we describe the methodology and
observations used for our analysis. Model-independent constraints on $H_0$ and $\Omega_{K}$ are
presented in Section~\ref{sec:constraint}. Finally, a brief summary and discussion are presented
in Section~\ref{sec:summary}.

\section{Methodology and Data}
\label{sec:method}

In a homogeneous and isotropic space, the spacetime geometry of the Universe can be described by the
FLRW metric
\begin{equation}
ds^{2}=-c^{2}dt^{2}+a^{2}(t)\left(\frac{dr^{2}}{1-Kr^{2}}+r^{2}d\Omega^{2}\right)\;,
\end{equation}
where $a(t)=1/(1+z)$ is the scale factor and the constant $K$ determines the spatial curvature.
The present value of the Hubble parameter $H(z)\equiv\dot{a}/a$ is labeled $H_0$. Let $D_A(z_l,z_s)$
denote the angular diameter distance of a source at redshift $z_s$ (corresponding to emission time
$t_s$) as observed at redshift $z_l$. Assuming that geometrical optics holds, the dimensionless comoving
distance $d(z_l,z_s)\equiv (1+z_s)H_0 D_A(z_l,z_s)/c$ (which is independent of $H_0$) is then given by
\begin{equation}
d(z_l,z_s)=\frac{1}{\sqrt{|\Omega_{K}|}}{\rm sinn}\left(\sqrt{|\Omega_{K}|}\int_{z_l}^{z_s}\frac{H_0}{H(z)}dz\right)\;,
\label{eq:dls}
\end{equation}
where $\Omega_{K}\equiv-K/H_0^{2}$ is the curvature parameter. Also, sinn is sinh when $\Omega_{K}>0$ and
sin when $\Omega_{K}<0$. For a flat Universe with $\Omega_{K}=0$, Equation~(\ref{eq:dls}) simplifies to
a linear function of the integral.
For convenience, we define $d(z)\equiv d(0,z)$, $d_l\equiv d(0,z_l)$, $d_s\equiv d(0,z_s)$, and $d_{ls}\equiv d(z_l,z_s)$.
If $d(z)$ is monotonic and $d^{'}(z)>0$, then these dimensionless distances in the FLRW frame are related via
the distance sum rule \citep{1993ppc..book.....P,2006ApJ...637..598B,2015PhRvL.115j1301R}:
\begin{equation}
d_{ls}=d_s \sqrt{1+\Omega_{K}d_{l}^{2}}-d_l \sqrt{1+\Omega_{K}d_{s}^{2}} \;.
\label{eq:DSR}
\end{equation}

It is worth noting that the FLRW metric can be ruled out if the derived $\Omega_{K}$ from the
combination of distances ($d_l$, $d_s$, and $d_{ls}$) are observationally found to be unequal
for any two pairs of ($z_l$, $z_s$). Furthermore, Equation~(\ref{eq:DSR}) can be rewritten as
\citep{2017ApJ...839...70L}
\begin{equation}
\frac{d_{ls}}{d_l d_s}=T(z_l)-T(z_s) \;,
\label{eq:DSR2}
\end{equation}
where
\begin{equation}
T(z)=\sqrt{1/d(z)^{2}+\Omega_{K}}\;,
\end{equation}
such that the distance $d(z)$ and the time-delay distance ratio $d_l d_s/d_{ls}$ (see below) are encoded.

In strong lensing, the measured time delay between two images of the source is related to both the
geometry of the Universe and the gravitational potential of the lens galaxy via the relation
\begin{equation}
\Delta t=\frac{D_{\Delta t}}{c}\Delta \phi \;,
\end{equation}
where $D_{\Delta t}$ is the time-delay distance and $\Delta \phi$ is the difference between the
Fermat potentials of the two images. The time-delay distance is the combination of three angular
angular diameter distances \citep{1964MNRAS.128..307R,1992grle.book.....S,2010ApJ...711..201S}:
\begin{equation}
D_{\Delta t}=\left(1+z_l\right)\frac{D_l D_s}{D_{ls}}=\frac{c}{H_0}\frac{d_l d_s}{d_{ls}} \;,
\label{eq:TDD}
\end{equation}
where subscripts ``$l$'' and ``$s$'' stand for lens and source, respectively. $D_{\Delta t}$ has units
of distance and is inversely proportional to $H_0$. Therefore, with measurements of $\Delta t$, $d_l$,
and $d_s$ and an accurate lens model to estimate $\Delta \phi$, we can directly determine $H_0$ and
$\Omega_{K}$ from Equations (\ref{eq:DSR2}) and (\ref{eq:TDD}) without involving any specific
cosmological model.

In this work, the time-delay distance ratio $d_l d_s/d_{ls}$ is extracted from strong gravitational
lensing, while the other two distances ($d_l$ and $d_s$) are obtained using the UV versus X-ray
luminosity correlation in quasars.

\subsection{Strong lensing data: time-delay distance ratios}

Recently, the H0LiCOW collaboration presented the latest measurements of $H_0$
from a combined sample of six strong lensing systems with measured time delays \citep{2019arXiv190704869W}.
The six lenses are B1608+656 \citep{2010ApJ...711..201S,2019Sci...365.1134J}, RXJ1131-1231
\citep{2013ApJ...766...70S,2014ApJ...788L..35S,2019MNRAS.490.1743C}, HE 0435-1223
\citep{2017MNRAS.465.4895W,2019MNRAS.490.1743C}, SDSS 1206+4332 \citep{2019MNRAS.484.4726B},
WFI2033-4723 \citep{2019arXiv190509338R}, and PG 1115+080 \citep{2019MNRAS.490.1743C}.
All lenses except B1608+656 were analyzed blindly with respect to the cosmological parameters.
We summarize the lens and source redshifts (i.e., $z_l$ and $z_s$), as well as the time-delay
distance constraint $D_{\Delta t}$, for each individual lens in Table~\ref{table1}. The posterior
distributions of the time-delay distances for the six lenses are available on the H0LiCOW
website.\footnote{http://www.h0licow.org} For the lens B1608+656, the time-delay distance
likelihood function was given as a skewed log-normal distribution:
\begin{equation}
\mathcal{L}_{D_{\Delta t}}=\frac{1}{\sqrt{2\pi}\left(x-\lambda_{D}\right)\sigma_{D}}\exp\left[-\frac{\left(\ln\left(x-\lambda_{D}\right)-\mu_{D}\right)^{2}}{2\sigma_{D}^{2}}\right]\;,
\end{equation}
with the parameters $\mu_{D}=7.0531$, $\sigma_{D}=0.22824$, and $\lambda_{D}=4000.0$, where
$x=D_{\Delta t}/$(1 Mpc). For the other five lenses, the posterior distributions of $D_{\Delta t}$
were released in the form of Monte Carlo Markov chains (MCMC). A kernel density estimator was used
to compute $\mathcal{L}_{D_{\Delta t}}$ from the chains \citep{2019arXiv190704869W}.
Very recently, the STRIDES collaboration presented the most precise measurement of
$H_0$ to date from a single time-delay lens DES J0408-5354 \citep{2020MNRAS.tmp.1051S}.
Table~\ref{table1} also lists the redshifts and the measured time-delay distance for this lens.
We use the time-delay distance posterior of DES J0408-5354 that was derived in \cite{2020MNRAS.tmp.1051S}.

Following \cite{2019PhRvL.123w1101C}, we also use constraints from the double-source-plane strong lens
SDSSJ0946+1006 \citep{2008ApJ...677.1046G}. The lensing galaxy in this system has a redshift of
$z_l=0.222$ and the redshift of the first source $s_1$ is $z_{s_1}=0.609$ \citep{2008ApJ...677.1046G},
while the redshift of the second source $s_2$ is taken to be at the peak of the photometric redshift
probability from \cite{2014MNRAS.443..969C}, i.e., $z_{s_2}=2.3$. The presence of two sources lensed
by the same foreground galaxy offers an accurate constraint on the cosmological scaling factor
\begin{equation}
\beta\equiv\frac{d_{ls_1}d_{s_2}}{d_{s_1}d_{ls_2}}=\frac{d_{ls_1}}{d_{l}d_{s_1}}\cdot\frac{d_{l}d_{s_2}}{d_{ls_2}}\;.
\end{equation}
Note that this ratio is sensitive to the curvature parameter $\Omega_{K}$, being independent of $H_0$.
In SDSSJ0946+1006, the cosmological scaling factor is constrained to be $\beta^{-1}=1.404\pm0.016$
\citep{2014MNRAS.443..969C}. That is, the posterior distribution of $\beta^{-1}$ is well approximated by
a Gaussian function centred at 1.404 with width $\sigma_{\beta^{-1}}=0.016$. The likelihood function
for $\beta^{-1}$ is then given by
\begin{equation}
\mathcal{L}_{\beta^{-1}}=\frac{1}{\sqrt{2\pi}\,\sigma_{\beta^{-1}}}\exp\left\{-\frac{\left[\beta^{-1}-
\frac{T\left(z_l\right)-T\left(z_{s2}\right)}{T\left(z_l\right)-T\left(z_{s1}\right)}
\right]^{2}}{2\sigma_{\beta^{-1}}^{2}}\right\}\;.
\end{equation}
%\left[T(z_l)-T(z_{s_2})\right]/\left[T(z_l)-T(z_{s_1})\right]

\subsection{Quasar data: the distances $d_{l}$ and $d_s$}
\label{subsec:quasar}

In order to obtain model-independent measurements of $H_0$ and $\Omega_{K}$ via Equation~(\ref{eq:DSR2}),
we also need to know the distance $d(z)$, which we here measure using the non-linear correlation between
the UV and X-ray luminosities of quasars. After refining their selection technique and flux measurements,
\cite{2019NatAs...3..272R} collected a final sample of 1598 quasars with reliable measurements of the
intrinsic UV and X-ray emissions. We use this high-quality quasar catalog covering the redshift range
$0.036<z<5.1$ for the analysis demonstrated in this paper. The non-linear luminosity relation of quasars,
$\log_{10}L_{\rm X}=\gamma \log_{10}L_{\rm UV}+\kappa$, can be re-expressed in terms of the measured UV
and X-ray fluxes, $F_{\rm UV}$ and $F_{\rm X}$, and the luminosity distance, $D_L$, at redshift $z$,
according to the expression
\begin{equation}
  \log_{10}F_{\rm X}=\kappa'+\gamma \log_{10}F_{\rm UV}+2\left(\gamma-1\right)\log_{10}D_{L}\;,
\label{eq:Fx}
\end{equation}
where $\kappa'$ is a parameter that subsumes the slope $\gamma$ and intercept $\kappa$, i.e.,
$\kappa'=\kappa+(\gamma-1)\log_{10}4\pi$. The luminosity distance $D_L$ can then be extracted from
the fluxes as a function of $\gamma$ and $\kappa'$. In practice, the parameter $\gamma$ can be
derived in a cosmology-independent way by directly fitting the relation between $F_{\rm X}$ and
$F_{\rm UV}$ with sub-samples in narrow redshift intervals. Using this approach, one can verify the
lack of evolution in this relation with redshift---an essential requirement for its implementation
to obtain quasar distances. \cite{2019NatAs...3..272R} showed that the parameter $\gamma$ does
not display any significant evolution; it appears to be a constant at all redshifts. Its average
value is $\gamma=0.633\pm0.002$. The redshift dependence of the scaling parameter $\kappa$ is
difficult to test without a solid physical explanation for the $L_X - L_{UV}$ relation
\citep{2015ApJ...815...33R,2019NatAs...3..272R}. Since its intrinsic value is still unknown,
we may instead regard $\kappa'$ as an arbitrary scaling factor. But following the treatment of
\cite{2015ApJ...815...33R,2019NatAs...3..272R}, we adopt their average value of $\gamma$
($=0.633\pm0.002$) to estimate a scaling parameter-dependent $D_L$ for each quasar
using Equation~(\ref{eq:Fx}).

With the distance-duality relation that holds true in any spacetime for any gravity theory
\citep{1933PMag...15..761E,2009GReGr..41..581E}, one can use the scaling parameter-dependent
luminosity distance of quasars to obtain the dimensionless comoving angular diameter distances
$d=H_{0}D_L/c(1+z)$. In principle, we need to select those quasars whose redshift matches that
of the lens and source in each system. It is difficult for this to be fulfilled for all discrete
observed events, however. There are always differences between the lensing redshifts and the
nearest quasars. This issue can be overcame by reconstructing a continuous distance function that
best approximates the discrete observed data using a polynomial fit \citep{2015PhRvL.115j1301R,2017ApJ...839...70L,2018ApJ...854..146L,2019PhRvL.123w1101C,2019PhRvD..99h3514L,2020ApJ...889..186Z}.
In our analysis, we construct the dimensionless distance function $d(z)$ in a cosmology
independent way by fitting a third-order polynomial with initial conditions $d(0)=0$ and $d'(0)=1$,
to the quasar data. This polynomial is expressed as
\begin{equation}
d(z)=z+a_{1}z^2+a_{2}z^3\;,
\label{eq:dz}
\end{equation}
where $a_1$ and $a_2$ are two free parameters that must be optimized along with the scaling
parameter $\kappa'$ and the intrinsic dispersion $\sigma_{\rm int}$ (see below).
We find that a third-order polynomial is flexible enough to fit the current data, while higher order
polynomials do not improve the goodness of fit, especially when taking into account the larger number
of free parameters.

The high-redshift quasar sample contains a significant intrinsic dispersion $\sigma_{\rm int}$,
which has to be treated as an additional free parameter \citep{2015ApJ...815...33R,2019NatAs...3..272R}.
Thus, the variance on each quasar is given by the quadratic sum of the measurement error of that quasar
($\sigma_{i}$) and $\sigma_{\rm int}$. This leads to the following formula of the likelihood function:
\begin{equation}
 \mathcal{L}_{\rm quasars} = \prod_{i=1}^{1598}\frac{1}{\sqrt{2\pi\left(\sigma_{i}^{2}+\sigma_{\rm int}^{2}\right)}}
 \exp\left[-\frac{\Delta_{i}^{2}}{2\left(\sigma_{i}^{2}+\sigma_{\rm int}^{2}\right)}\right]\;,
\label{eq:likelihood-QSO}
\end{equation}
where
\begin{eqnarray}
\Delta_{i}&=& \nonumber \log_{10}(F_{\rm X})_{i}-\gamma \log_{10}(F_{\rm UV})_{i} \\
          & & -\kappa'-2\left(\gamma-1\right)\log_{10}\left[\frac{c}{H_0}\left(1+z_i\right)d(z_i)\right]\;.
\end{eqnarray}
In this likelihood estimation, there is a degeneracy between the Hubble constant $H_0$ and $\kappa'$.
We therefore adopt a fiducial $H_0=70$ km $\rm s^{-1}$ $\rm Mpc^{-1}$ for the sake of
optimizing $\kappa'$. Another choice of $H_0$ would require a corresponding re-scaling of this
optimized $\kappa'$.

\begin{figure*}
%\vskip-0.1in
\centerline{\includegraphics[keepaspectratio,clip,width=0.9\textwidth]{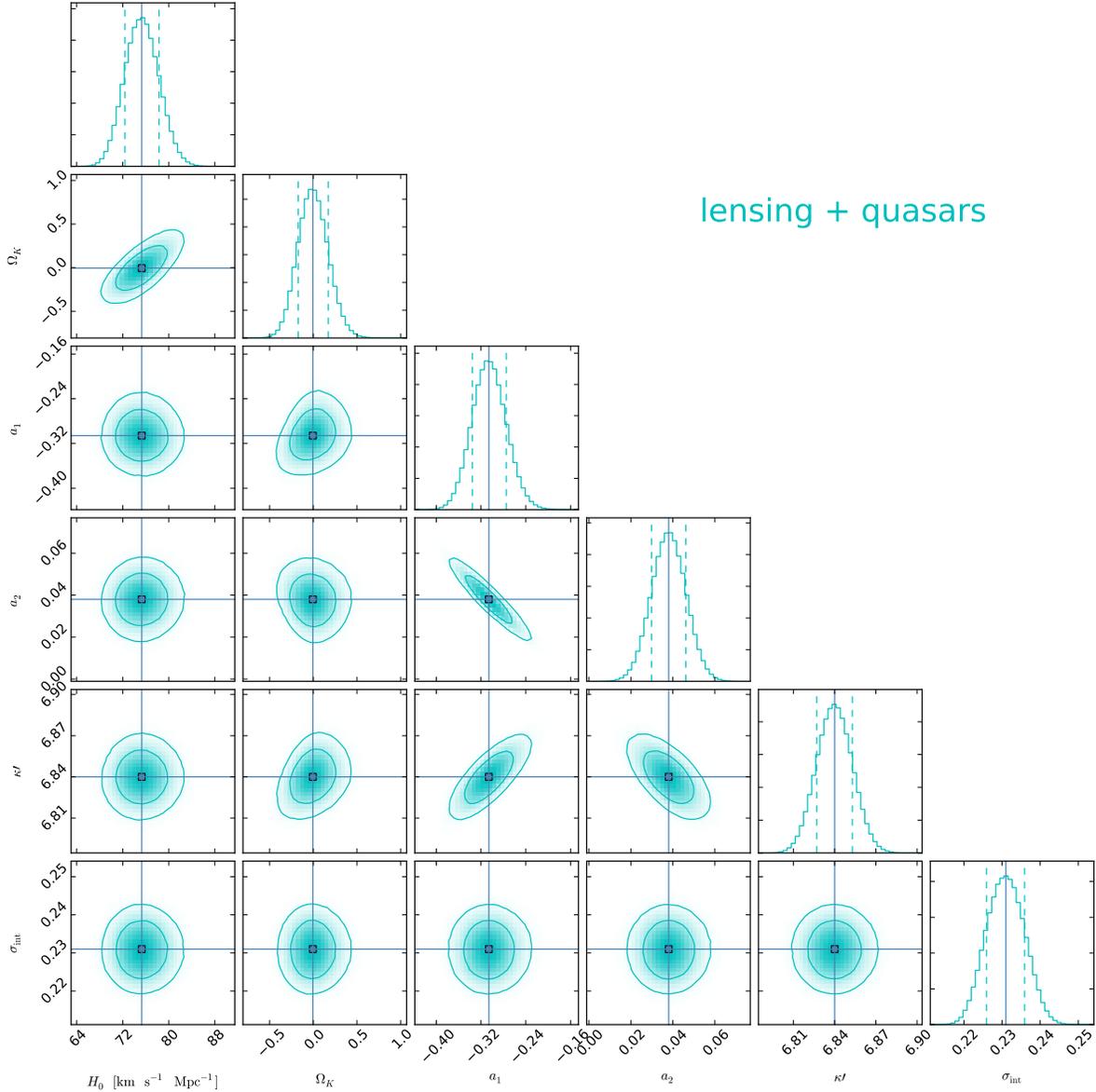}}
\vskip-0.1in
\caption{1D and 2D marginalized probability distributions with $1\sigma$ and $2\sigma$
confidence contours for the parameters $H_0$, $\Omega_{K}$, $a_1$, $a_2$, $\kappa'$, and $\sigma_{\rm int}$
constrained by the strong lensing
systems and quasars. The vertical solid lines represent the medium values, and the vertical dashed
lines enclose the 68\% credible region.}
\label{f1}
\end{figure*}

\section{Model-independent Constraints on $H_0$ and $\Omega_K$}
\label{sec:constraint}

The quantities $d(z)$, $H_0$, and $\Omega_{K}$ are fitted to the strong lensing and quasar data
simultaneously using the Python MCMC module EMCEE \citep{2013PASP..125..306F}. The final
log-likelihood sampled by EMCEE is a sum of the separate likelihoods of the time delay lenses,
double-source-plane strong lens, and high-redshift quasars:
\begin{equation}
\ln\left(\mathcal{L}_{\rm tot}\right) = \ln\left(\mathcal{L}_{D_{\Delta t}}\right) + \ln\left(\mathcal{L}_{\beta^{-1}}\right)
   + \ln\left(\mathcal{L}_{\rm quasars}\right)\;.
\end{equation}
The third-order polynomial has two free parameters ($a_1$ and $a_2$). The scaling parameter $\kappa'$ and
the intrinsic dispersion $\sigma_{\rm int}$ enter into the quasar likelihood as two nuisance parameters.
In addition, $d_l d_s/d_{ls}$ given by Equation~(\ref{eq:DSR2}) involves the curvature parameter $\Omega_{K}$
and the time-delay distance given by Equation~(\ref{eq:TDD}) depends on $H_0$, making it six free
parameters in total.

\begin{table*}
\centering \caption{Constraints on All Parameters with Various Choices of Data}
\begin{tabular}{cccccccc}
\hline
\hline
 Data &  $H_0$  &  $\Omega_{K}$  &  $a_1$  &  $a_2$   &  $\kappa'$  &  $\sigma_{\rm int}$  & $M_B$ \\
  &  (km $\rm s^{-1}$ $\rm Mpc^{-1}$)  &    &    &     &     &     &   \\
\hline
lensing+quasars & $75.3^{+3.0}_{-2.9}$ & $-0.01^{+0.18}_{-0.17}$ & $-0.306^{+0.031}_{-0.030}$ & $0.038^{+0.008}_{-0.008}$  & $6.840^{+0.013}_{-0.013}$ & $0.231^{+0.005}_{-0.005}$ & -- \\
lensing+quasars & $75.3^{+1.9}_{-1.9}$ & 0 (fixed) & $-0.301^{+0.029}_{-0.028}$ & $0.037^{+0.008}_{-0.008}$  & $6.841^{+0.012}_{-0.012}$ & $0.231^{+0.005}_{-0.005}$ & -- \\
lensing+SNe Ia & $75.9^{+3.1}_{-3.1}$ & $0.16^{+0.22}_{-0.20}$ & $-0.259^{+0.017}_{-0.017}$ & $0.032^{+0.012}_{-0.012}$ & -- & -- & $-19.344^{+0.011}_{-0.011}$ \\
lensing+SNe Ia & $74.3^{+1.9}_{-1.9}$ & 0 (fixed) & $-0.252^{+0.015}_{-0.015}$ & $0.025^{+0.008}_{-0.008}$ & -- & -- & $-19.346^{+0.010}_{-0.010}$ \\
lensing+quasars+SNe Ia & $75.3^{+3.0}_{-2.9}$ & $0.05^{+0.16}_{-0.14}$ & $-0.260^{+0.012}_{-0.012}$ & $0.027^{+0.004}_{-0.004}$ & $6.856^{+0.008}_{-0.008}$ & $0.231^{+0.005}_{-0.005}$ & $-19.341^{+0.009}_{-0.009}$  \\
lensing+quasars+SNe Ia & $74.5^{+1.7}_{-1.7}$ & 0 (fixed) & $-0.259^{+0.012}_{-0.012}$ & $0.026^{+0.004}_{-0.004}$ & $6.856^{+0.008}_{-0.008}$ & $0.231^{+0.005}_{-0.005}$ & $-19.341^{+0.009}_{-0.009}$  \\
\hline
\end{tabular}
\label{table2}
\end{table*}

\begin{figure}
\vskip-0.2in
\centerline{\includegraphics[keepaspectratio,clip,width=0.6\textwidth]{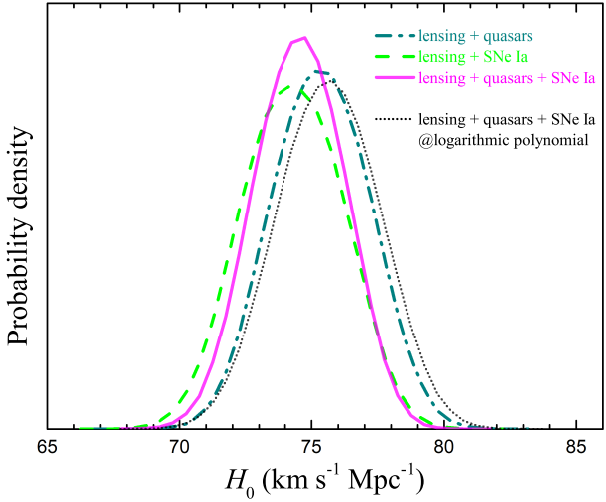}}
\vskip-0.2in
\caption{1D marginalized probability distributions of the Hubble constant $H_0$ in a flat Universe
constrained from the lensing + quasar data (dot-dashed curve), the lensing + SN Ia data (dashed curve),
and the combined lensing + quasar + SN Ia data (solid curve), respectively, using a linear polynomial
function. The dotted curve corresponds to the analysis of the lensing + quasar + SN Ia data
using a logarithmic polynomial function.}
\label{f2}
\end{figure}

The 1D marginalized probability distributions and 2D regions with $1\sigma$ and $2\sigma$
contours corresponding to these six parameters, constrained by the lensing + quasar data, are displayed in Figure~\ref{f1}. These contours show that, at the 68\% confidence level, the median values and
the 16th and 84th percentiles are $H_0=75.3^{+3.0}_{-2.9}$ km $\rm s^{-1}$ $\rm Mpc^{-1}$,
$\Omega_{K}=-0.01^{+0.18}_{-0.17}$, $a_1=-0.306^{+0.031}_{-0.030}$, $a_2=0.038^{+0.008}_{-0.008}$,
$\kappa'=6.840^{+0.013}_{-0.013}$, and $\sigma_{\rm int}=0.231^{+0.005}_{-0.005}$. If we instead
assume zero spatial curvature, the marginalized probability distribution for $H_0$ is shown in
Figure~\ref{f2} (dot-dashed curve). This model-independent constraint yields $H_0=75.3^{+1.9}_{-1.9}$ km
$\rm s^{-1}$ $\rm Mpc^{-1}$. The corresponding results for the lensing + quasar data are summarized
in lines 1 and 2 of Table~\ref{table2} for a non-flat and flat Universe. The comparison between
these two cases indicates that the nuisance parameters ($a_1$, $a_2$, $\kappa'$, and $\sigma_{\rm int}$)
have little effect on the cosmological parameters.
%By marginalizing the polynomial coefficients and the parameters characterizing the quasar
%luminosity relation, we obtain the 1D and 2D probability distributions with $1-2\sigma$
%confidence regions for $H_0$ and $\Omega_{K}$ displayed in Figure~\ref{f1}.

For this analysis, we have used the average value ($0.633$) of the slope $\gamma$ in the quasar
luminosity relation (Eqn.~\ref{eq:Fx}), estimated within narrow redshift bins, as described in
Section~\ref{subsec:quasar}. \cite{2019MNRAS.489..517M} tested the dependence of this slope
on the choice of cosmological model, and found that it appears to be very weakly dependent on
the expansion rate as well, showing that $\gamma$ falls within the relatively narrow range
of $(0.626,0.640)$ for three diverse formulations of the luminosity distance $D_L$. To
investigate how sensitive our results on $H_0$ and $\Omega_{K}$ are on the choice of
$\gamma$ within this range, we also perform two parallel comparative analyses of the
lensing + quasar data using $\gamma=0.626$ and $0.640$. For the latter, we find
$H_0=75.5^{+3.0}_{-2.9}$ km $\rm s^{-1}$ $\rm Mpc^{-1}$ and $\Omega_{K}=0.04^{+0.17}_{-0.16}$.
For the former, we get $H_0=75.1^{+3.0}_{-2.9}$ km $\rm s^{-1}$ $\rm Mpc^{-1}$ and
$\Omega_{K}=-0.05^{+0.18}_{-0.17}$. The values of $H_0$ and $\Omega_{K}$ change only slightly
as $\gamma$ is varied, with the largest variation lying within $\approx 0.2\sigma$ of the optimized
values, well within the uncertainty. So there is no concern regarding a possibly large dependence
on $\gamma$. For the rest of the paper, we shall therefore adopt the average value
$\gamma=0.633$ estimated by \cite{2019NatAs...3..272R} in their study of narrow redshift bins.

The SNe Ia are often adopted as distance indicators for providing the distance $d(z)$
on the right side of Equation~(\ref{eq:DSR2}) (see, e.g., \citealt{2015PhRvL.115j1301R,2018ApJ...854..146L,
2019ApJ...886L..23L,2020arXiv200210605L,2019PhRvL.123w1101C}). To compare our results with previous
work, we therefore also carry out our model-independent analysis using a combination of data that
includes the latest Pantheon SN Ia observations, and see if one can further constrain the comoving
distance $d(z)$. \cite{2018ApJ...859..101S} recently released the largest combined sample of SNe Ia
referred to as Pantheon, consisting of 1048 SNe Ia in the redshift range of $0.01 < z < 2.3$.
The distance modulus of a Type Ia SN can be determined using the Spectral Adaptive
Light curve Template 2 (SALT2) light-curve fit parameters,
based on the formula $\mu=m_{B}-M_{B}+ \alpha X_{1}-\eta \mathcal{C}+\Delta_{M}+\Delta_{B}$,
where $m_B$ is the observed \emph{B}-band apparent magnitude, $X_{1}$ is the light-curve stretch
factor, $\mathcal{C}$ is the color, $\Delta_{M}$ is a distance correction based on the host galaxy
mass, and $\Delta_{B}$ denotes a distance correction based on predicted biases from simulations.
Furthermore, $\alpha$ and $\eta$ are nuisance coefficients of the luminosity-stretch and luminosity-color
relations, respectively, and $M_{B}$ is another nuisance parameter that describes the absolute
\emph{B}-band magnitude of a fiducial SN.

\begin{figure*}
%\vskip-0.1in
\centerline{\includegraphics[keepaspectratio,clip,width=0.8\textwidth]{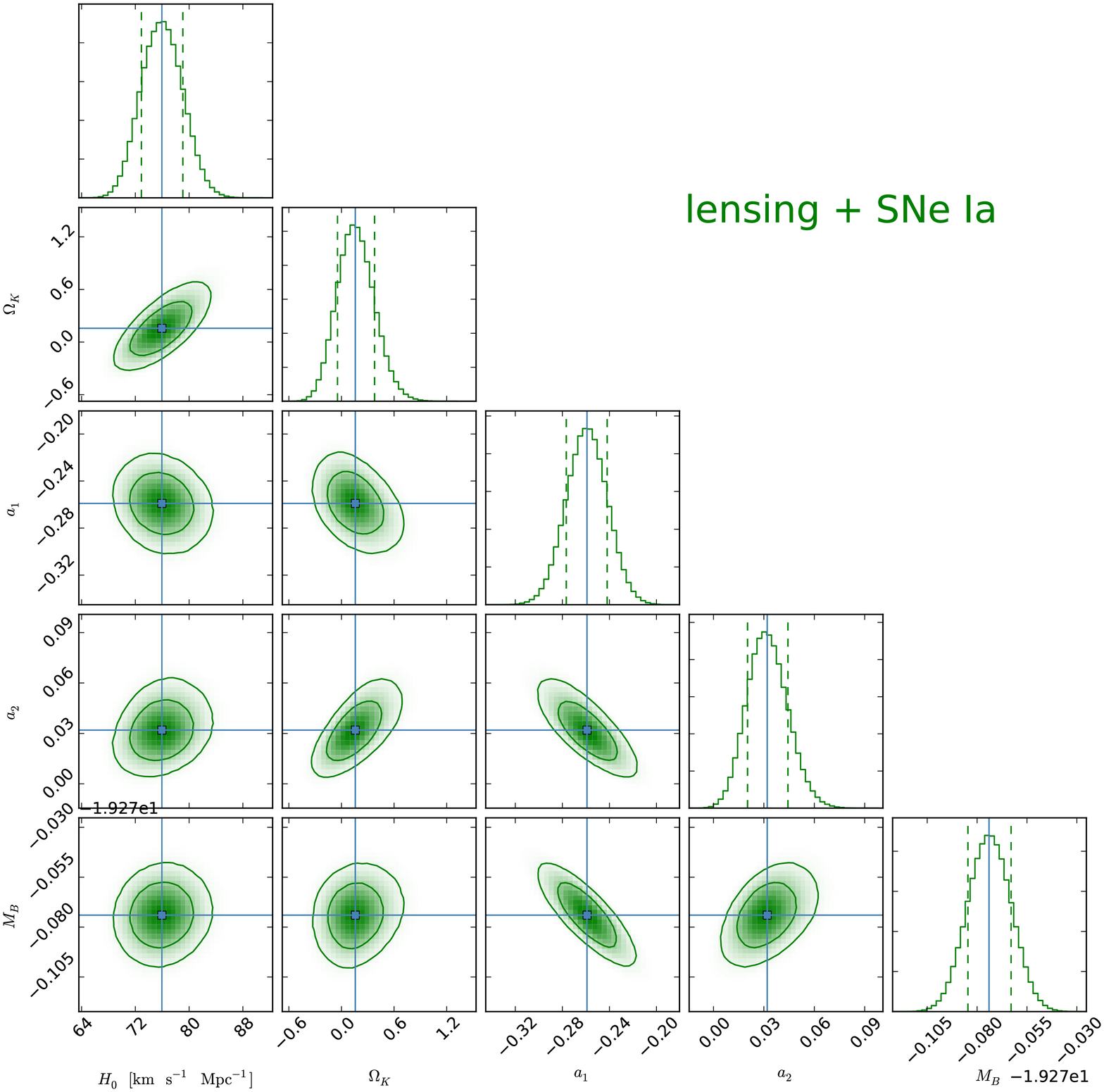}}
\vskip-0.1in
\caption{Same as Figure~\ref{f1}, but now showing the constraints for the parameters
$H_0$, $\Omega_{K}$, $a_1$, $a_2$, and $M_B$ based on the analysis of the
combined strong lensing and Pantheon SN Ia data only.}
\label{f3}
\end{figure*}

In general, the two nuisance parameters $\alpha$ and $\eta$ should be optimized simultaneously
with the cosmological parameters for each specific cosmological model. In this case, the derived
SN distances are model-dependent. To dodge this problem, \cite{2017ApJ...836...56K} introduced
the BEAMS with Bias Corrections (BBC) method to correct those expected biases and simultaneously
fit for the $\alpha$ and $\eta$ parameters. This method relies on the approach proposed by
\cite{2011ApJ...740...72M}, but involves extensive simulations for correcting the SALT2 light-curve
fitter. The BBC fit creates a bin-averaged Hubble diagram of SNe Ia, and then the coefficients
$\alpha$ and $\eta$ are inferred by fitting to a reference cosmological model.  The reference
cosmology is supposed to reproduce the local shape of the Hubble diagram within each redshift bin.
If there are sufficient redshift bins, the fitted coefficients $\alpha$ and $\eta$ will converge
to consistent values, which are independent of the reference cosmology \citep{2011ApJ...740...72M}.
With the BBC method, \cite{2018ApJ...859..101S} reported the corrected apparent magnitudes
$m_{\rm corr}=\mu+M_{B}$ for all the Pantheon SNe. Therefore, we just need to subtract $M_{B}$ from
$m_{\rm corr}$ to obtain the observed distance modulus $\mu$. For the SN data set, the
uncertainties are given by a covariance matrix $\textbf{C}$ (including both statistical and
systematic uncertainties). Given a vector of distance residuals of the SN sample that may be
defined as $\Delta \bf{\hat{\mu}}=\bf{\hat{\mu}}-\bf{\hat{\mu}_{\rm model}}$,
where $\bf{\hat{\mu}}$ ($\bf{\hat{\mu}_{\rm model}}$) is the observed (model) vector of distance moduli,
the likelihood of the model fit is expressed as
\begin{equation}
-2 \ln\left(\mathcal{L}_{\rm SN}\right) = \Delta \bf{\hat{\mu}}^{\emph{T}} \cdot \textbf{C}^{-1} \cdot \Delta \bf{\hat{\mu}}\;.
\end{equation}
Here the model vector $\bf{\hat{\mu}_{\rm model}}$ is determined by
$\mu_{{\rm model},i}=5\log_{10}[D_{L}(\mathcal{P},z_i)/{\rm 10\; pc}]=5\log_{10}[(1+z_i)d(\mathcal{P},z_i)]+M_{H_0}$,
where $M_{H_0}=-5\log_{10}({\rm 10\; pc}\;H_0/c)$ and $\mathcal{P}$ denotes the model parameters.
Given the degeneracy with the absolute magnitude $M_{B}$, the value of $M_{H_0}$ is arbitrary and we fix it to
$M_{H_0}=43.16$ (corresponding to the aforementioned fiducial $H_0=70$ km $\rm s^{-1}$ $\rm Mpc^{-1}$).

\begin{figure*}
%\vskip-0.1in
\centerline{\includegraphics[keepaspectratio,clip,width=1.0\textwidth]{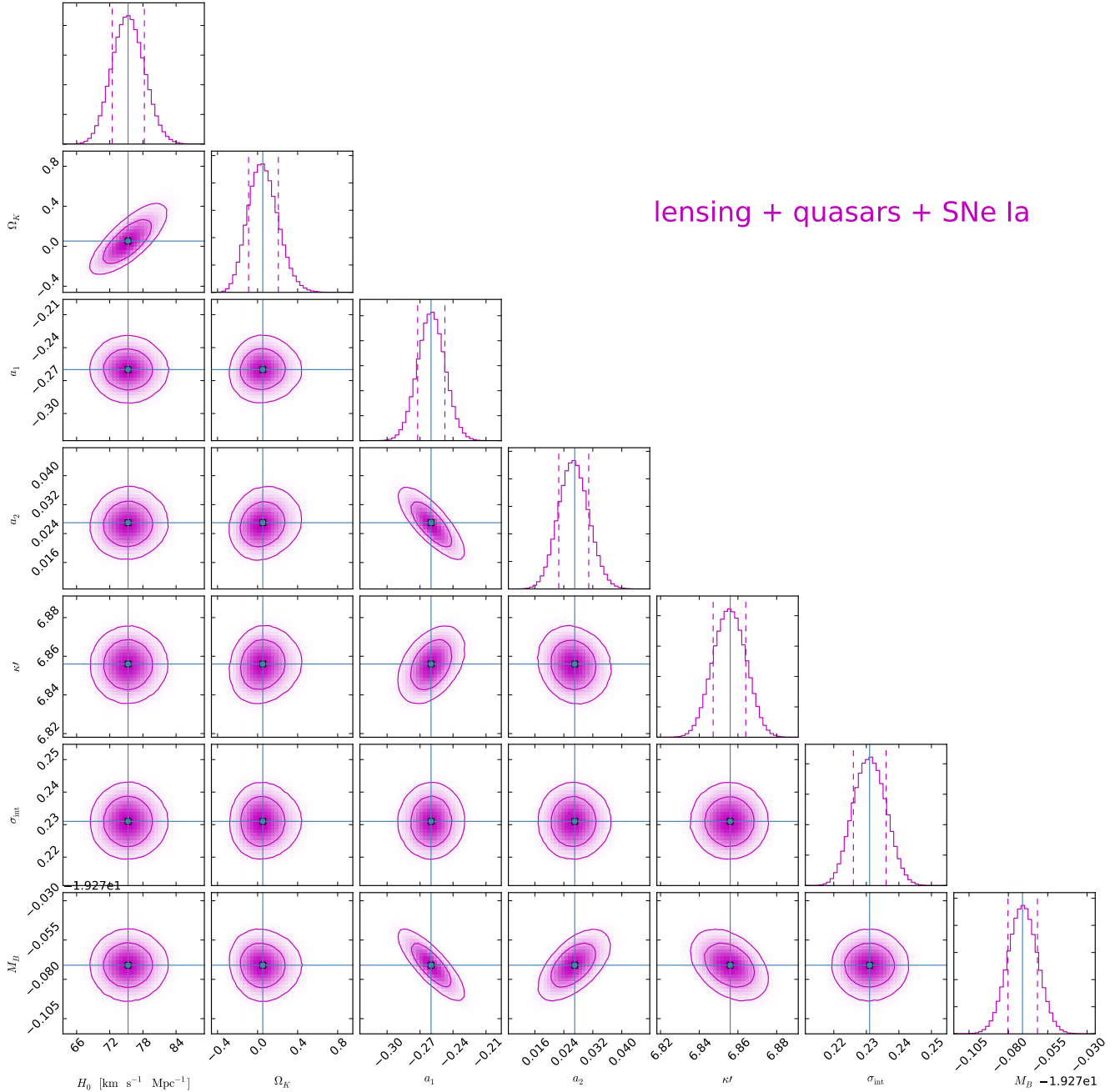}}
\vskip-0.1in
\caption{Same as Figure~\ref{f1}, but now showing the constraints for the parameters
$H_0$, $\Omega_{K}$, $a_1$, $a_2$, $\kappa'$, $\sigma_{\rm int}$, and $M_B$ based on the analysis
of the combined strong lensing, quasar, and Pantheon SN Ia data.}
\label{f4}
\end{figure*}

We first analyze the lensing + Pantheon SN Ia data. In this case, \
the log-likelihood sampled by EMCEE becomes
\begin{equation}
\ln\left(\mathcal{L}_{\rm tot}\right) = \ln\left(\mathcal{L}_{D_{\Delta t}}\right) + \ln\left(\mathcal{L}_{\beta^{-1}}\right) + \ln\left(\mathcal{L}_{\rm SN}\right)\;.
\end{equation}
There are five free parameters, including the Hubble constant $H_0$, the curvature parameter $\Omega_{K}$,
the two polynomial coefficients ($a_1$ and $a_2$), and the SN absolute magnitude $M_{B}$.
These parameters are constrained to be $H_0=75.9^{+3.1}_{-3.1}$ km $\rm s^{-1}$ $\rm Mpc^{-1}$,
$\Omega_{K}=0.16^{+0.22}_{-0.20}$, $a_1=-0.259^{+0.017}_{-0.017}$, $a_2=0.032^{+0.012}_{-0.012}$,
and $M_{B}=-19.344^{+0.011}_{-0.011}$, which are presented in Figure~\ref{f3} and Table~\ref{table2}.
If a prior of flatness (i.e., $\Omega_{K}=0$) is adopted, the marginalized $H_0$ constraint is
$H_0=74.3^{+1.9}_{-1.9}$ km $\rm s^{-1}$ $\rm Mpc^{-1}$ (see dashed curve in Figure~\ref{f2}).
The comparison between lines 1 and 3 of Table~\ref{table2} suggests that the constraint precisions
of $H_0$ and $\Omega_{K}$ obtained using the lensing + quasar data are slightly better than those from
the lensing + SN Ia data.

We also carry out this type of analysis using the combined strong lensing + quasar +
Pantheon SN Ia data sets. The final log-likelihood sampled by EMCEE now becomes
\begin{equation}
\ln\left(\mathcal{L}_{\rm tot}\right) = \ln\left(\mathcal{L}_{D_{\Delta t}}\right) + \ln\left(\mathcal{L}_{\beta^{-1}}\right)
   + \ln\left(\mathcal{L}_{\rm quasars}\right) + \ln\left(\mathcal{L}_{\rm SN}\right)\;.
\end{equation}
In this case, the free parameters are the Hubble constant $H_0$, the curvature parameter $\Omega_{K}$,
the two polynomial coefficients ($a_1$ and $a_2$), the parameters characterizing the quasar luminosity
relation ($\kappa'$ and $\sigma_{\rm int}$), and the SN absolute magnitude $M_{B}$.
As shown in Figure~\ref{f4} and Table~\ref{table2}, the marginalized distributions give
$H_0=75.3^{+3.0}_{-2.9}$ km $\rm s^{-1}$ $\rm Mpc^{-1}$, $\Omega_{K}=0.05^{+0.16}_{-0.14}$,
$a_1=-0.260^{+0.012}_{-0.012}$, $a_2=0.027^{+0.004}_{-0.004}$, $\kappa'=6.856^{+0.008}_{-0.008}$,
$\sigma_{\rm int}=0.231^{+0.005}_{-0.005}$, and $M_{B}=-19.341^{+0.009}_{-0.009}$.
We see that, compared to the results obtained using solely the strong lensing and quasar observations,
somewhat more precise constraints may be achieved for $H_0$ and $\Omega_{K}$ by also including
the SNe Ia data. The marginalized constraint on $H_0$ assuming a flat Universe is shown in
Figure~\ref{f2} (solid curve), and is given by $H_0=74.5^{+1.7}_{-1.7}$ km $\rm s^{-1}$ $\rm Mpc^{-1}$.

\section{Summary and Discussion}
\label{sec:summary}

Based on the sum rule of distances along null geodesics of the FLRW metric, we can obtain model-independent
constraints on both the Hubble constant $H_0$ and spatial curvature $\Omega_{K}$ by confronting observations
of strong lensing time delays with SN Ia luminosity distances. In this paper, aiming to mitigate the
redshift limitation of using solely SNe Ia as distance indicators, we have proposed using high-$z$
quasars to provide the distances required by the sum rule. Combining the time delay measurements
of six H0LiCOW lenses and one STRIDES lens with the known UV versus X-ray luminosity correlation of
1598 quasars, we have simultaneously placed limits on $H_0$ and $\Omega_{K}$ without assuming any
specific cosmological model. This analysis suggests that the curvature parameter is constrained to be
$\Omega_{K}=-0.01^{+0.18}_{-0.17}$, consistent with a flat Universe. Meanwhile, the optimized
Hubble constant is $H_0=75.3^{+3.0}_{-2.9}$ km $\rm s^{-1}$ $\rm Mpc^{-1}$. If instead we assume
a spatially flat universe, we find $H_0=75.3^{+1.9}_{-1.9}$ km $\rm s^{-1}$ $\rm Mpc^{-1}$,
representing a precision of 2.5\%, in good agreement with the measurement of $H_0$ using
SNe Ia calibrated by the local distance ladder which, however, is in $4.0\sigma$ tension
with the value inferred by {\it Planck} from CMB measurements. These model-independent
results are fully consistent with the Hubble constant inferred previously from the H0LiCOW data
assuming a flat $\Lambda$CDM model \citep{2019arXiv190704869W}.

We also carried out this type of analysis using the combined strong lensing + Pantheon SN Ia
data sets and the combined strong lensing + quasar + Pantheon SN Ia data sets, respectively.
For the former, we found that the model-independent constraints are $H_0=75.9^{+3.1}_{-3.1}$ km $\rm s^{-1}$ $\rm Mpc^{-1}$ and $\Omega_{K}=0.16^{+0.22}_{-0.20}$, which are slightly worse than the constraint
precisions obtained with the lensing + quasar data. For the latter,
we found that the constraints on $H_0$ and $\Omega_{K}$ may be improved
somewhat, yielding $H_0=75.3^{+3.0}_{-2.9}$ km $\rm s^{-1}$ $\rm Mpc^{-1}$ and
$\Omega_{K}=0.05^{+0.16}_{-0.14}$. And if a flat Universe is assumed as a prior, one derives the
optimized value $H_0=74.3^{+1.9}_{-1.9}$ km $\rm s^{-1}$ $\rm Mpc^{-1}$ (representing
a 2.6\% precision measurement) for the former case, and $H_0=74.5^{+1.7}_{-1.7}$ km $\rm s^{-1}$ $\rm Mpc^{-1}$ (representing a 2.3\% precision measurement) for the latter case.

Previously, \cite{2019arXiv190704869W} obtained $H_0=74.4^{+2.1}_{-2.3}$ km $\rm s^{-1}$ $\rm Mpc^{-1}$
and $\Omega_{K}=0.26^{+0.17}_{-0.25}$ by analyzing six time-delay lenses in the non-flat $\Lambda$CDM model.
\cite{2019PhRvL.123w1101C} used time delay lens and SNe Ia to obtain model-independent constraints of
$H_0=75.7^{+4.5}_{-4.4}$ km $\rm s^{-1}$ $\rm Mpc^{-1}$ and $\Omega_{K}=0.12^{+0.27}_{-0.25}$ by implementing
a polynomial fitting to the supernova luminosity distances. And by using Gaussian processes to
extract the supernova distances, \cite{2020arXiv200210605L} obtained model-independent determinations of
$H_0=77.3^{+2.2}_{-3.0}$ km $\rm s^{-1}$ $\rm Mpc^{-1}$ and $\Omega_{K}=0.33^{+0.12}_{-0.19}$ from
strong lensing and SN Ia data. Comparing our results to these previous constraints, it is quite
apparent that our method is at least competitive with these other approaches. Most importantly, our
method offers a new model-independent way of simultaneously constraining both $H_0$ and $\Omega_{K}$.

\begin{figure*}
%\vskip-0.1in
\centerline{\includegraphics[keepaspectratio,clip,width=1.0\textwidth]{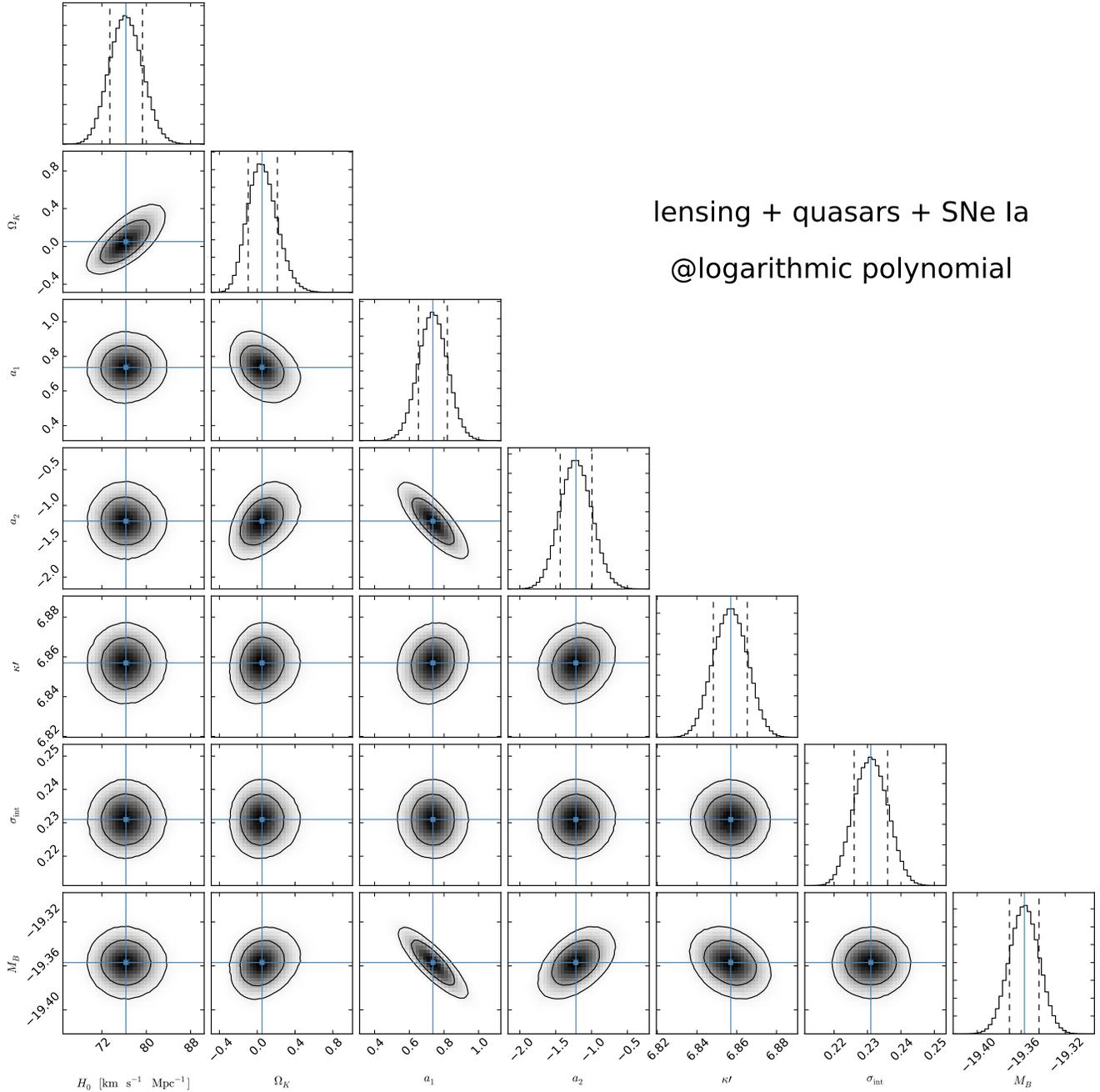}}
\vskip-0.1in
\caption{Same as Figure~\ref{f4}, but now showing the constraints based on the analysis
of the combined lensing + quasar + SN Ia data using a logarithmic polynomial function.}
\label{f5}
\end{figure*}

Finally, we considered whether our choice of parametrization for the dimensionless distance $d(z)$
(as a linear polynomial function; see Equation~\ref{eq:dz}) might be affecting the results.
To test the dependence of the outcome on the functional form of $d(z)$, we also carried out
a parallel comparative analysis of the lensing + quasar + SN Ia data using a logarithmic polynomial function, i.e., $d(z)=\ln(10)[\log_{10}(1+z)+a_{1}\log_{10}^{2}(1+z)+a_{2}\log_{10}^{3}(1+z)]$, like that used
by \cite{2019NatAs...3..272R}, and the results are summarized in Figure~\ref{f5}.
Using the logarithmic polynomial fit, we found that the constraints
are $H_0=76.3^{+3.0}_{-2.9}$ km $\rm s^{-1}$ $\rm Mpc^{-1}$, $\Omega_{K}=0.05^{+0.16}_{-0.15}$,
$a_1=0.736^{+0.084}_{-0.083}$, $a_2=-1.218^{+0.224}_{-0.219}$, $\kappa'=6.857^{+0.008}_{-0.009}$,
$\sigma_{\rm int}=0.231^{+0.005}_{-0.005}$, and $M_{B}=-19.357^{+0.013}_{-0.013}$.
Assuming a zero spatial curvature, we got $H_0=75.6^{+1.9}_{-1.9}$ km $\rm s^{-1}$ $\rm Mpc^{-1}$
(see dotted curve in Figure~\ref{f2}; representing a precision of 2.5\%),
$a_1=0.747^{+0.075}_{-0.075}$, $a_2=-1.248^{+0.193}_{-0.189}$, $\kappa'=6.857^{+0.008}_{-0.008}$,
$\sigma_{\rm int}=0.231^{+0.005}_{-0.005}$, and $M_{B}=-19.358^{+0.013}_{-0.013}$.
The differences between these optimized cosmological parameters and those obtained with
the linear polynomial fit are well within the $1\sigma$ errors. Thus, the adoption of
a different functional form for $d(z)$ has only a minimal influence on these results.

\acknowledgments
We would like to thank the anonymous referee for constructive comments
which helped improve our work. We are also grateful to Anowar Jaman Shajib for
sharing the distance posteriors of the lens DES J0408-5354, and to Shen-Shi Du for his kind help.
This work is partially supported by the National Natural Science Foundation of China
(grant Nos. 11673068, 11725314, and U1831122), the Youth Innovation Promotion
Association (2017366), the Key Research Program of Frontier Sciences (grant Nos. QYZDB-SSW-SYS005
and ZDBS-LY-7014), the Strategic Priority Research Program ``Multi-waveband gravitational wave universe''
(grant No. XDB23000000) of Chinese Academy of Sciences,
and the Guangxi Key Laboratory for Relativistic Astrophysics.

%\bibliographystyle{apj}
%\bibliography{ms}

\end{document}